
\input phyzzx
\FRONTPAGE
\line{\hfill BROWN-HET-922}
\line{\hfill August 1993}
\vskip1.5truein
\titlestyle{{SOME STATISTICS FOR MEASURING LARGE-SCALE
STRUCTURE}}
\bigskip
\author{Robert H. BRANDENBERGER, David M. KAPLAN\footnote{1)}{Present
address: Physics Department, University of California, Santa Barbara,
CA 93106, USA} and Stephen A. RAMSEY\footnote{2)}{Present address:
Department of Physics and Astronomy, University of Maryland, College
Park, MD 20742, USA}}
\centerline{{\it Department of Physics}}
\centerline{{\it Brown University, Providence, RI 02912, USA}}
\bigskip
\abstract
Good statistics for measuring large-scale structure in the Universe
must be able to distinguish between different models of structure
formation.  In this paper, two and three dimensional ``counts in cell"
statistics and a new ``discrete genus statistic" are applied to toy
versions of several popular theories of structure formation:  random
phase cold dark matter model, cosmic string models, and global texture
scenario.  All three statistics appear quite promising in terms of
differentiating between the models.
\endpage
\noindent \underbar{1. Introduction}
\par
Recent redshift surveys extending to greater than 100h$^{-1}$ Mpc seem,
upon visual inspection, to be dominated by voids, sheets and
filaments$^{1,2)}$.  Are these structures real and are they
significant?  Are they consistent with currently popular models of
structure formation?  In order to answer these questions, we need
statistical methods.  Since most of the currently popular theories of
structure formation predict a similar (namely scale invariant)
spectrum of density perturbations$^{3)}$, we require statistical
measures which can pick out the phase information which distinguishes
between the models.

More specifically, we are interested in statistics which can clearly
distinguish between the two most popular classes of theories: models
based on random phase fluctuations produced during inflation ({\it
e.g.}, the cold dark matter (CDM) model) on one hand and topological
defect models on the other.  A good statistic must also be able to
differentiate between the various topological defect models -- as
representative examples we pick the global texture scenario, a model
based on cosmic string wakes, and a filament model.

In this paper we discuss three promising statistics: two and three
dimensional counts in cell (CIC) statistics$^{4)}$, and a discrete
genus statistic.  We apply these statistics to our toy models of
structure formation and conclude that for sufficiently small
observational error bars the statistics will be able to clearly
differentiate between the models.

In Section 2 we define our toy models.  In Section 3 we define the
``discrete genus statistic" and apply it to our toy models.
In Section 4, we investigate
the two dimensional and three dimensional CIC statistics.  We compare
the results of the two dimensional CIC statistic with the results for
two slices of the CFA2 survey$^{1)}$.   The final
section contains a discussion of the results and ideas for future
work.

This paper is based on senior theses by D.K.$^{5)}$ and S.R.$^{6)}$.

\vskip.5cm \noindent \underbar{2. Toy Models}
\par
The purpose of this paper is primarily to study the effectiveness of
the statistics considered here at distinguishing different models of
structure formation.  At this stage we are not yet attempting to confirm
or rule out concrete models.  Hence, we will apply the statistics to
{\it toy models} of structure formation.  These models are designed to mimic
key features of specific theories of structure formation, in
particular the distinctive non-gaussian and topological aspects.

We consider five models: a model in which galaxies are randomly
distributed throughout the sample volume (Poisson model), a CDM model
(without nonlinearities taken into account), a cosmic string wake
model, a cosmic string filament model, and a global texture model.

In order to study the dependence of the statistical measures on
topology (rather than number density), we chose all topological defect
models to contain the same number of structures, one per Hubble volume
at $t_{eq}$, the time of equal matter and radiation.  All structures
in a given model have the same mass, with the total mass chosen to
give a spatially flat Universe.

Taking the structures to have the same size corresponds to a severe
truncation of the power spectrum of the actual topological defect
models.  The justification for this truncation comes from the fact
that structures produced at $t_{eq}$ are dominant in both the global
texture models and in cosmic string models in which the dark matter is
hot.  We will come back to this point below.

The numerical simulations produce cubes of data whose side length is
200 Mpc and which contains 222,400 galaxies, chosen such that the number of
galaxies per unit volume agrees roughly with the number density of the
CFA2 survey$^{1)}$.

In the texture toy model, spherical balls of galaxies with Gaussian
radial density function were placed randomly in the sample volume.
The standard deviation of the Gaussian was taken to be 9 Mpc.

This toy model should provide a rough approximation for what happens
in the actual texture model$^{7,8)}$  In this theory, density
perturbations are caused by contracting topologically nontrivial
scalar field configurations.  There is a fixed probability $p$ per
Hubble volume that at any time $t$ a nontrivial configuration will
become smaller than the Hubble radius and start to contract at
relativistic speeds, leading to a roughly spherical density
perturbation.  Hence, a fixed number $p$ of textures per Hubble volume
per expansion time are created.  Those produced before $t_{eq}$ are
washed out by pressure, those produced after $t_{eq}$ have less time
to grow by gravitational instability.  Hence, the most prominent
texture induced perturbations are those laid down at $t_{eq}$.

The cosmic string model of galaxy formation$^{9)}$ still has many
uncertain aspects.  It is known that the network of cosmic strings
approaches a scale invariant distribution$^{10)}$, {\it i.e.}, the
distribution of strings looks statistically the same at all times
provided all lengths are scaled by the Hubble radius.  The cosmic
string ensemble consists of a network of infinite strings with
curvature radius comparable to time $t$, and a distribution of loops
with radius smaller than $t$.  Recent numerical simulations$^{11)}$
agree that the loops are subdominant.  However, there is no agreement
on the small scale structure on long strings.

If long strings are straight on small scales, they will form planar
density perturbations called wakes$^{12)}$ (see {\it e.g.}, Ref. 13
for a recent review of structure formation in the cosmic string model)
of planar dimensions $t \times vt$, where $v$ is the velocity of the
string in its normal plane.  However, if there is small scale
structure on the long strings, these strings will move slowly and will
exert a local gravitational force on the surrounding matter, leading
to the formation of filaments$^{14)}$.  Because of this uncertainty in
the string model we consider two cosmic string toy models, a ``wake
model" and a ``filament model."  They correspond to the two extreme
cosmic string scenarios.

In the wake model, rectangular prisms of length and width 40 Mpc and
thickness 2 Mpc were placed randomly in the sample volume (subject to
the constraint that they lie entirely in the sample volume).  The
thickness corresponds to the thickness of the nonlinear region around
the wake for a cosmic string model with hot dark matter and a mass per
unit length $\mu$ given by $G \mu = 10^{-6}$, $G$ being Newton's
constant$^{15)}$.  This value of $\mu$ is the preferred value based on
large-scale structure analyses$^{16)}$ and on the COBE cosmic
microwave anisotropy results$^{17)}$.  Note that for $h=0.5$, the
planar dimensions of the wake correspond to the Hubble radius at
$t_{eq}$.

In the filament model, cylinders of length 60 Mpc and radius 4.1 Mpc
were placed randomly in the sample volume.  Galaxies were placed at
random in the cylinders, as they were in the wake model.

The linear CDM model was constructed by starting from the power
spectrum$^{18)}$
$$
| \delta (\underline{k}) |^2 = {Ak\over{(1 + \beta k + \omega k^{1.5} + \gamma
k^2)^2 }} \eqno\eq
$$
with
$$
\eqalign{ \beta & = 1.7 (\Omega_0 h^2)^{-1} {\rm Mpc} \cr
\omega & = 9 (\Omega_0 h^2)^{-1.5} {\rm Mpc}^{1.5} \cr
\gamma & = 1 (\Omega_0 h^2)^{-2} {\rm Mpc}^2 \cr} \, , \eqno\eq
$$
Fourier transforming to position space, and laying down galaxies
according to the position space density distribution.  The transition
from Fourier space to position space was done by taking the lowest
50$^3$ Fourier modes (corresponding to the sample volume) in the first
octant of Fourier space, choosing random phases for all of these modes,
by evaluating the Fourier transform at 50$^3$ cell centers
$\underline{x}_{ijk}$ in position space, by calculating the number of
galaxies in cell $(ijk)$ according to $\rho (\underline{x}_{ijk})$ and
by laying down the galaxies at random in the cell (for details see
Ref. 6).

\vskip.5cm \noindent \underbar{3. Discrete Genus Statistic}
\par
The first statistic we investigate is a variant (developed in Ref. 5)
of the genus statistic which was proposed in 1986 by Gott et
al.$^{19)}$ as a method of gaining direct information about the
topology of the galaxy distribution.

For a compact surface $S$ in $\Re^3$, the genus is defined as
$$
g = {\rm (\# ~ of ~ holes}) - {\rm (\# ~ of ~ disconnected ~ components)} + 1
\,
. \eqno\eq
$$
By the Gauss-Bonnet theorem the genus can be computed as a surface
integral of the Gaussian curvature $k$:
$$
g = - {1\over{4 \pi}} \int\limits_S k dA \, . \eqno\eq
$$

Given a smooth density distribution $\rho (\underline{x})$, the genus
statistic is defined as the curve $g (\rho)$, where $g (\rho)$ is the
genus of the surface $\rho (\underline{x}) = \rho$.  For a random
phase density field, the genus curve can be calculated
analytically$^{19)}$:
$$
g (\nu) = N (1 - \nu^2) \exp (-\nu^2 /2) \, , \eqno\eq
$$
where $\nu$ is the number of standard deviations from the mean density,
and $N$ is a constant which depends on the power spectrum.  Note that
$g (\nu)$ is peaked at $\nu = 0$ and is symmetric.  For non-Gaussian
models we expect a shift in the peak position and a deviation from
symmetry about $\nu = 0$.

The usual method$^{19)}$ of applying the genus statistic to a
distribution of galaxies is to construct a smooth density field by
smearing each galaxy with a Gaussian distribution
$$
W(r) = {1\over{\pi^{3/2} \lambda^3}} \exp (-r^2 / \lambda^2) \, ,
\eqno\eq
$$
where $r$ is the smoothing length.

The choice of $\lambda$ is critical.  $\lambda$ must be large enough
such that the density distribution inside structures connect, but
small enough such that the topology of the dominant structures is not
lost.  The results depend crucially on $\lambda$, and this is a big
disadvantage of the statistic.

To avoid the above problem, we use a ``discrete genus
statistic"$^{5)}$.  Given a volume limited redshift survey (or a
simulated galaxy distribution), we divide the volume into cells of
size smaller than that of the structures we are interested in probing
but large enough such that the counts in cell are not dominated by
shot noise.  In our simulations we chose a cell size of 8 Mpc.

Consider the polygonal surface $S(n)$ which is the boundary of the
complex of cells each of which contains greater than or equal to $n$ galaxies.
The genus $g (n)$ of this surface is
$$
g (n) = 1 - {1\over 2} (V-E+F) \eqno\eq
$$
where $V,E$ and $F$ are the number of vertices, edges and faces
respectively.

The curve $g(n)$ is the discrete genus curve.  The surface $S(n)$ can
be regarded as the surface with galaxy density $n$/(cell volume).
Hence, we can plot $g$ as a function of the galaxy number density.

The results of the numerical simulations are shown in Fig. 1.  With
exception of the CDM model, the data is the average of 20 independent
simulations of the model. The statistical error bars are smaller than the
symbol sizes. The CDM model results come from a single
realization.

The most important conclusion we can draw from this investigation is
that the discrete genus statistic is a very powerful discriminant
between different models of structure formation.  The genus curves for
all topological defect models are highly asymmetrical about the mean
number density, whereas the Poisson and CDM models are symmetrical.  The width
of the genus curve for the CDM model is larger than that for the Poisson
model which reflects the degree of clustering in the simulation.  The
difference in the peak density is due to a slightly different
normalization of the models.

For the wake the genus curve is positive.  This is due to the many
holes between the interconnected network of wakes.  In contrast, the
genus curve for the texture model is overwhelmingly negative since the
distribution of galaxies is clumpy (no holes and many disconnected
components).  The curve for the filament model lies between the two
extreme cases.

\vskip.5cm \noindent \underbar{4. Counts in Cell Statistics}

The counts in cell statistics$^{4)}$ are very simple.  The sample
volume is divided into cells of equal volume.  For each integer $n$,
the number $f(n)$ of occurrences of cells with $n$ galaxies is
determined.  The graph of $f(n)$ as a function of $n$ is the counts in
cell statistic (CIC). Counts in cell statistics have been studied extensively
by Saslaw and collaborators$^{20)}$, and more recently by Coles and
Plionis$^{21)}$ for the Lick galaxy catalog, by Coles et al.$^{22)}$ for CDM
models, by Kaiser et al.$^{23)}$ for IRAS galaxies, by Weinberg and
Cole$^{24)}$ and by de Lapparent et al.$^{25)}$ in the context of defining a
percolation statistic.

For our three dimensional simulations, it is straightforward to
evaluate the CIC.  We divide the simulation box into 50$^3$ cells,
each on the average containing about two galaxies.  Simulations with a
smaller number of cells showed more noise whereas the range of $n$
values with $f(n) \neq 0$ was too small for a greater number of cells.

The results of the simulations are shown in Figs. 2 and 3.  As in
Section 3, the results are averages over 20 simulations, except for
the CDM model for which only a single realization was considered.  In
the region of $n$ values plotted in Fig. 2, the one sigma statistical
error bars are of the size of the symbols, as is seen from the
individual plots of Fig. 3.

The most obvious conclusion is that the three dimensional CIC
statistic can well discriminate between our toy models.  The CIC curve
for the texture model has the longest tail, a reflection of the dense
clusters of galaxies it contains.  The length of the tail of the CIC
curve decreases as the dimension of the structures of the model
increases.  This allows a clear distinction between the filament and
wake models.  All topological defect toy models considered here are
more strongly clustered than the CDM model and hence have longer tails
of the CIC statistic.

We can also consider two dimensional CIC statistics.  They are
constructed such that a comparison with data from the CFA2$^{1)}$
redshift survey is possible.  Slices of data designed to resemble the
CFA2 slices were extracted from cubes of simulated data by generating
random orientations for the slices and selecting all galaxies in the
cube within the angular $(120^\circ \times 6^\circ)$ and radial
$(100{\rm h}^{-1}{\rm Mpc})$ bounds of the slice.  The slices were
divided into 35$^2$ cells of equal volume.  Less cells per side
reduced the resolution and generated a lot of noise, whereas more
cells per side caused a significant shortening of the CIC curves.
Considering cells of equal area instead of volume would weight nearby and far
away galaxies differently.  This explains our choices.

In order to calculate the CIC statistic for the CFA data we must correct for
the apparent magnitude limitation of the data set. This was done in two steps.
First, a volume-limited subsample of the data with radial extent $100
h^{-1}$Mpc
was used. The volume thus selected includes most of the interesting structure
in each of the CFA slices, but is small enough such that selection effects can
be reliably corrected.

Second, the number of galaxies in each cell was multiplied by a selection
function $f(r)$, $r$ being the distance of the cell from us, which corrects for
the deficiency of galaxies. This function $f(r)$ can be determined from the
Schechter luminosity function$^{26)}$
$$
\varphi(L) dL = \varphi^* ({L \over {L^*}})^{\alpha} exp(- L/L^*) d({L \over
{L^*}}),\eqno\eq
$$
where $\varphi(L) dL$ is the number density of galaxies with luminosity in the
interval $[L, L + dL]$, and $\alpha, \varphi^*$ and $L^*$ are parameters
determined from the data. The CFA data gives$^{25)}$
$$\alpha = -1.1,$$
$$\varphi^* = 0.020 h^3 Mpc^{-3},\eqno\eq$$
$$M^* = -19.2,$$
where $M^*$ is the absolute magnitude corresponding to $L^*$.

The number density of galaxies which can be seen at a distance $r$ given the
apparent magnitude cutoff in the data is
$$ \phi(r) = {\int_{L(r)}^{\infty}} \varphi(L) dL, \eqno\eq$$
where $L(r)$ is the absolute magnitude which at distance $r$ corresponds to the
apparent magnitude cutoff. The selection function $f(r)$ is
$$ f(r) = {{\phi(r_0)} \over {\phi(r)}},\eqno\eq$$
where $r_0$ is a suitably chosen reference distance ($20 h^{-1}$Mpc in our
case).

The results of our simulations are shown in Figs. 4 and 5.  In Fig.
4 the results are compared to the average of two CFA2 slices.  The
error bars of the individual CFA2 data were determined from the
uncertainty in the positions due to peculiar velocities.  The statistical error
bars of the numerical simulations are shown in Fig. 5.

The tendency of the two dimensional CIC curves is the same as for
three dimensions: the texture curve has the longest tail, followed by
the wake and filamentary models.  All three defect models give rise to
CIC curves with longer tails than the CDM model.  However, the
observational error bars are sufficiently large such that only the
Poisson model is convincingly ruled out.  A $\chi^2$ analysis shows
that the string filament model fits the data best, significantly
better than the CDM model$^{5)}$.

At this stage, however, it is premature to draw conclusions about the
validity of the various models of structure formation.  The toy models
are too naive to allow any such conclusion.  The main lesson is that
both two and three dimensional CIC statistics are good ways to analyze
large-scale structure data and confront theory with observations.

\vskip.5cm \noindent \underbar{5. Discussion}

We have studied the applicability of a discrete genus statistic and
two and three dimensional counts in cell statistics to distinguish the
predictions of different models of structure formation.  Most theories
predict a similar power spectrum of density perturbations, and hence a
good statistic must be able to pick out the non-random phases which
differentiate between the models.

We conclude that our three statistics give large differences when
applied to toy models of structure formation.  Topological defect
models give rise to long tails in counts in cell statistics, the tail
length increasing as the dimension of the prominent structure
decreases.  The discrete genus statistic is very sensitive to the
topology of large-scale structure and shows a large difference between
the texture and cosmic string wake toy models.

We have applied the statistics to five toy models of structure
formation: Poisson, CDM, global texture, cosmic string wakes and
cosmic string filaments.  The models are constructed to capture the
important topological and statistical properties of the ``real" models
on scales larger than the horizon at $t_{eq}$.  On smaller scales, the
topological defect toy models are too rough to give a good
approximation to the actual models.

In this paper we have only compared one set of data, namely two slices
of the CFA2 redshfit survey, with the toy models.  In future work we
plan to analyze more data.  We also plan to construct more realistic
toy models for topological defect models which have the correct power
spectrum on all scales and are normalized to agree with the CMB
anisotropies measured on large angular scales by the COBE-DMR
experiment$^{27)}$.  It will then be realistic to perform a detailed
statistical comparison between toy models and observations.

\ack

This work is supported in part by DOE Grant DE-FG02-91ER40688, Task A.
We are grateful to John Huchra for providing us with the CFA2 data
files.  One of us (R.B.) wishes to thank the Aspen Center for Physics
for hospitality during the 1993 workshop on Large-Scale Structure after COBE.
\endpage
\REF\one{V. de Lapparent, M. Geller and J. Huchra, {\it Ap. J.
(Lett.)} {\bf 302}, L1 (1986); \nextline
M. Geller and J. Huchra, {\it Science} {\bf 246}, 897 (1989).}
\REF\two{R. Kirshner, A. Oemler, P. Schechter and S. Shechtman, {\it
Ap. J. (Lett.)} {\bf 248}, L57 (1981); \nextline
Ya. Zel'dovich, J. Einasto and S. Shandarin, {\it Nature} {\bf 300},
407 (1982); \nextline
J. Oort, {\it Ann. Rev. Astr. Astrophys.} {\bf 21}, 373 (1983);
\nextline
R. Tully, {\it Ap. J.} {\bf 257}, 381 (1982); \nextline
S. Gregory, L. Thomson and W. Tifft, {\it Ap. J.} {\bf 243}, 411
(1980); \nextline
J. Einasto, M. Joeveer and E. Saar, {\it Mon. Not. R. astron. Soc.}
{\bf 193}, 353 (1980); \nextline
R. Giovanelli and M. Haynes, {\it Astron. J.} {\bf 87}, 1355 (1982);
\nextline
D. Batuski and J. Burns, {\it Ap. J.} {\bf 299}, 5 (1985).}
\REF\three{R. Brandenberger, ``Topological Defect Models of Structure
Formation after the COBE Discovery of CMB Anisotropies," in proc. of
the Int. School of Astrophysics `D. Chalonge,' 6-13 Sept. 1992, ed. N.
Sanchez (World Scientific, Singapore, 1993).}
\REF\four{W. Saslaw, {\it Ap. J.} {\bf 297}, 49 (1985).}
\REF\five{D.M. Kaplan, ``Statistics for Measuring Large-Scale
Structure," Senior thesis, Brown University (1993).}
\REF\six{S.A. Ramsey, ``A Statistic for Measuring Large-Scale
Structure," Senior thesis, Brown Unversity (1992).}
\REF\seven{N. Turok, {\it Phys. Rev. Lett.} {\bf 63}, 2625 (1989).}
\REF\eight{N. Turok, {\it Phys. Scripta} {\bf T36}, 135 (1991), and
references therein.}
\REF\nine{Ya. Zel'dovich, {\it Mon. Not. R. astron. Soc.} {\bf 192},
663 (1980); \nextline
A. Vilenkin, {\it Phys. Rev. Lett.} {\bf 46}, 1169 (1981).}
\REF\ten{A. Vilenkin, {\it Phys. Rev. Lett.} {\bf 121}, 263 (1985);
\nextline
A. Albrecht and N. Turok, {\it Phys. Rev. Lett.} {\bf 54}, 1868
(1985).}
\REF\eleven{D. Bennett and F. Bouchet, {\it Phys. Rev. Lett.} {\bf 60}, 257
(1988); \nextline
B. Allen and E.P.S. Shellard, {\it Phys. Rev. Lett.} {\bf 64}, 119
(1990); \nextline
A. Albrecht and N. Turok, {\it Phys. Rev.} {\bf D40}, 973 (1989).}
\REF\twelve{J. Silk and A. Vilenkin, {\it Phys. Rev. Lett.} {\bf 53},
1700 (1984); \nextline
T. Vachaspati, {\it Phys. Rev. Lett.} {\bf 57}, 1655 (1986); \nextline
A. Stebbins et al., {\it Ap. J.} {\bf 322}, 1 (1987).}
\REF\thirteen{R. Brandenberger, {\it Phys. Scripta} {\bf T36}, 114
(1991).}
\REF\fourteen{T. Vachaspati and A. Vilenkin, {\it Phys. Rev. Lett.}
{\bf 67}, 1057 (1991); \nextline
D. Vollick, {\it Phys. Rev.} {\bf D45}, 1884 (1992).}
\REF\fifteen{L. Perivolaropoulos, R. Brandenberger and A. Stebbins,
{\it Phys. Rev.} {\bf D41}, 1764 (1990).}
\REF\sixteen{N. Turok and R. Brandenberger, {\it Phys. Rev.} {\bf
D33}, 2175 (1986); \nextline
A. Stebbins, {\it Ap. J. (Lett.)} {\bf 303}, L21 (1986); \nextline
H. Sato, {\it Prog. Theor. Phys.} {\bf 75}, 1342 (1986).}
\REF\seventeen{D. Bennett, A. Stebbins and F. Bouchet, {\it Ap. J.
(Lett.)} {\bf 399}, L5 (1992); \nextline
L. Perivolaropoulos, {\it Phys. Lett.} {\bf 298B}, 305 (1993).}
\REF\eighteen{P.J.E. Peebles, {\it Ap. J.} {\bf 263}, L1 (1983);
\nextline
M. Davis, G. Efstathiou, C. Frenk and S. White, {\it Ap. J.} {\bf
292}, 371 (1985).}
\REF\nineteen{J. Gott, A. Melott and M. Dickinson, {\it Ap. J.} {\bf
306}, 341 (1986); \nextline
J. Gott, D. Weinberg and A. Melott, {\it Ap. J.} {\bf 319}, 1
(1987).}
\REF\twenty{W. Saslaw, {\it Ap. J.} {\bf 341}, 588 (1989);\nextline
A. Hamilton, W. Saslaw and T. Thuan, {\it Ap. J.} {\bf 297}, 37
(1985);\nextline
P. Crane and W. Saslaw, {\it Ap. J.} {\bf 301}, 1 (1986);\nextline
M. Itoh, S. Inagaki and W. Saslaw, {\it Ap. J} {\bf 331}, 45 (1988);\nextline
W. Saslaw and P. Crane, {\it Ap. J.} {\bf 380}, 315 (1991).}
\REF\twentyone{P. Coles and M. Plionis, {\it Mon. Not. R. astron. Soc.} {\bf
250}, 75 (1991).}
\REF\twentytwo{P. Coles et al., {\it Mon. Not. R. astron. Soc.} {\bf 260}, 572
(1993).}
\REF\twentythree{N. Kaiser et al., {\it Mon. Not. R. astron. Soc.} {\bf 252}, 1
(1991).}
\REF\twentyfour{D. Weinberg and S. Cole, {\it Mon. Not. R. astron. Soc.} {\bf
259}, 652 (1992).}
\REF\twentyfive{V. de Lapparent, M. Geller and J. Huchra, {\it Ap. J.} {\bf
369}, 273 (1991).}
\REF\twentsix{P. Schechter, {\it Ap. J.} {\bf 203}, 297 (1976).}
\REF\twentyseven{G. Smoot et al., {\it Ap. J. (Lett.)} {\bf 396}, L1
(1992).}
\refout
\endpage
\centerline{\bf Figure Captions}
\bigskip
\item{} {\bf Figure 1:} The discrete genus statistic evaluated for the four
models of structure formation considered in the text, and compared to the
results for a Poisson distribution of galaxies.
\item{} {\bf Figure 2:} 3-d counts in cell statistic evaluated for the four toy
models and for a Poisson distribution of galaxies.
\item{} {\bf Figure 3:} 3-d counts in cell statistic for the four toy models.
One sigma statistical error bars are shown for the wake, filament and texture
models.
\item{} {\bf Figure 4:} 2-d counts in cell for the filament, texture, Poisson
and inflation-based CDM models, compared to the mean $f(n)$ for two CFA slices.
\item{} {\bf Figure 5:} 2-d counts in cell (including statistical error bars)
for wake, filament, texture and CDM models.
\end